\documentclass[a4paper,fleqn,usenatbib]{mnras}

\usepackage{newtxtext,newtxmath}

\usepackage[T1]{fontenc}
\usepackage{ae,aecompl}
\usepackage[normalem]{ulem}

\usepackage{graphicx}	
\usepackage{amsmath}	

\newcommand{\dust}{{\rm d}}
\newcommand{\gas}{{\rm g}}
\newcommand{\kmsmpc}{\kms\;{\rm Mpc}^{-1}}

\newcommand{\hkpc}{h^{-1}{\rm kpc}}
\newcommand{\hmpc}{h^{-1}{\rm Mpc}}

\newcommand{\kms}{\;{\rm km}\,{\rm s}^{-1}}

\newcommand{\msolar}{\;{\rm M}_{\odot}}

\newcommand{\gad}{{\sc Gadget-3}}
\newcommand{\gizmo}{{\sc Gizmo}}

\newcommand{\simba}{{\sc Simba}}

\newcommand{\cm}{\,{\rm cm}}

\newcommand{\diff}{{\rm d}}
\newcommand\nH{n_{\rm H}}
\newcommand\mH{m_{\rm H}}

\usepackage{color}

\usepackage{comment}

\title[]{The Origin of the Dust Extinction Curve in Milky Way-like Galaxies}

\author[Q. Li, Narayanan, Torrey, Dav\'e \& Vogelsberger]{Qi Li$^{1}$\thanks{E-mail:
 pg3552@ufl.edu}, Desika Narayanan$^{1,2}$, Paul Torrey$^{1}$, Romeel Dav\'e$^{3,4,5}$ \& Mark Vogelsberger$^{6}$\\$^{1}$Department of Astronomy,
  University of Florida, 211 Bryant Space Science Center, Gainesville,
 FL, 32611, USA\\$^{2}$Cosmic Dawn Center (DAWN), Niels Bohr Institute, University of Copenhagen, Juliane Maries vej 30, DK-2100 Copenhagen, Denmark\\$^{3}$Institute for Astronomy, Royal Observatory Edinburgh, EH9 3HJ, UK\\$^{4}$University of the Western Cape, Bellville, Cape Town 7535, South Africa\\$^{5}$South African Astronomical Observatory, Cape Town 7925, South Africa\\$^6$Department of Physics, Kavli Institute for Astrophysics and Space Research, Massachusetts Institute of Technology, Cambridge, MA 02139, USA}%

\begin{document}

\date{}

\pagerange{\pageref{firstpage}--\pageref{lastpage}} \pubyear{2015}

\maketitle

\begin{abstract}
We develop a cosmological model for the evolution of dust grains in galaxies with a distribution of sizes in order to understand the origin of the Milky Way dust extinction curve.  Our model considers the formation of active dust in evolved stars, growth by accretion and coagulation, and destruction processes via shattering, sputtering, and astration in the ISM of galaxies over cosmic time.  Our main results follow.  Galaxies in our cosmological model with masses comparable to the Milky Way's at $z\sim 0$ exhibit a diverse range of extinction laws, though with slopes and bump strengths comparable to the range observed in the Galaxy.  The progenitors of the Milky Way have steeper slopes, and only flatten to slopes comparable to the Galaxy at $z \approx 1$.  This owes to increased grain growth rates at late times/in high-metallicity environments driving up the ratio of large to small grains, with a secondary dependence on the graphite to silicate ratio evolution. The UV bump strengths depend primarily on the graphite to silicate ratio, and remain broadly constant in MW-like galaxies between $z=3$ and $z=0$, though show slight variability.  Our models span comparable regions of bump-slope space as sightlines in the Galaxy do, though there is a lack of clear relationship between the model slopes and bump strengths owing to small scale fluctuations in the bump strength.  Our models naturally produce slopes for some non-Milky Way analogs as steep as those of the LMC and SMC in metal poor galaxies, though notably the bump strengths are, on average, too large when comparing to the Magellanic clouds. This owes to the fact that we evolve the grain size distributions of graphites and silicates simultaneously, which is an oversimplification.  Our model provides a novel framework to study the origins and variations of dust extinction curves in galaxies over cosmic time.

\end{abstract}

\section{Introduction}
\label{section:introduction}
The extinction of photons by dust in the interstellar medium (ISM) of galaxies is the amount of radiation lost on an individual sightline owing to either absorption or scattering away from the line of sight (see the recent reviews by \citet{galliano2018} and \citet{Salim2020})\footnote{This is to be distinguished by "attenuation" which quantifies the net loss of light from many unresolved sightlines, and therefore includes the impact of both scattering back into the line of sight, as well as the contribution to the observed signal by unobscured stars.}.  Generally, the wavelength-dependent nature of extinction is characterized by a family of curves or "laws" that rise toward shorter wavelengths, with a bump in extinction near $2175$\AA\ that is often referred to as the UV bump \citep{Stecher1965}.  Quantifying the shape and normalization of extinction laws is critical for de-reddening UV-to-near infrared observations to properly estimate physical quantities such as star formation rates and stellar masses.  Generally, the extinction law represents a convolution of the dust grain size distribution in the ISM and the ratio of the extinction cross section to the geometric cross section of those grains.

The best observational constraints of extinction curves come from nearby sources, owing to the need for well-resolved sightlines.  A fundamental requirement for these measurements is {\it a priori} knowledge of the shape of the unreddened spectra or spectral energy distribution (SED).  In the Milky Way, the most common method is to use what is known as the pair method:  here, the SED of an observed star is compared against one that is dust-free, and of a similar spectral type \citep{Stebbins1939,Stecher1965,Fitzpatrick1986}, though some studies have instead employed theoretical stellar atmosphere models as the dust free references \citep{Fitzpatrick2007}. 

More recently, a number of studies have employed large-survey statistical approaches to deriving extinction laws.  As an example, \citet{Peek2010} developed a method monikered "standard crayons" which used passive galaxies as standard background sources to measure the reddening at high Galactic latitudes.  This method was expanded by \citet{Berry2012} and \citet{Schlafly2014,Schlafly2016}, who used SED fitting to derive the intrinsic SEDs of stars from the large scale surveys (e.g. Sloan digital Sky Survey (SDSS), Two-Micron All Sky Survey (2MASS), Pan-STARRS1, and APOGEE).   Other large scale surveys to map the dust extinction law in the Galaxy include \citet{Schlafly2010}, \citet{Schlafly2011} and \citet{Wang2019}.

The seminal studies by \citet{Fitzpatrick1988} and \citet[][]{Cardelli1989} computed the extinction law within the Milky Way over tens of individual sightlines, and determined that even within an individual galaxy there is substantial dispersion in both the slopes and UV bump strength of the curves.  \citeauthor{Cardelli1989} further inferred that the slopes of these curves between the $V$-band and UV wavelengths were correlated with $R_{\rm V}^{-1} \equiv \left(A_{\rm B}/A_{\rm V}-1\right)$, though the scatter is significant.  In this paper, we will regularly compare to both the range of curves observed by \citet{Cardelli1989}, as well as the average \citet{Fitzpatrick2007} curve.

While the extinction law in the Galaxy may follow a self-similar family of curves (though this is unclear; \citet{Salim2020}), this does not extend to the Magellanic clouds, which demonstrate (on average) as a function of decreasing metallicity, steeper slopes with reduced UV bump strengths \citep{Clayton1985,Fitzpatrick1985}.  While, like the Galaxy, significant sightline-dependent variation exists \citep[e.g.][]{Prevot1984,Pei1992,Gordon1998,Gordon2003}, on average, the LMC is steeper than the Milky Way, and the SMC is steeper yet.  Outside of the Galaxy and Clouds, extinction law constraints in M31 \citep{Dong2014}, and galaxies outside of the Local Group \citep[using back lights such as other galaxies, GRBs or quasars;][]{White1992,York2006,Stratta2007,Holwerda2009,Zafar2011,Zafar2018} have demonstrated strong variations in both their slopes and 2175\AA\ UV bump strengths.  The origin of variations in extinction law slopes and bump strengths in galaxies is currently unknown, and an area in which theoretical models can provide some insight. 

Broadly, models for extinction curves fall into two categories: synthesis and numerical (see \citet{Salim2020} for a review).  In the former category, the goal is to develop a theory that simultaneously models the grain size distribution and composition of dust in the face of numerous observational constraints that include (but are not limited to) the observed wavelength-dependent extinction curve, polarization signatures, and broadband infrared emission features.  \citet{Mathis1977} (hereafter, MRN) developed a seminal model in which they found a power-law size distribution and combination of graphite-silicate grains performed well in fitting the existing Galactic extinction constraints (this model was subsequently expanded on substantially by \citealt{Draine1984} and \citealt{Laor1993}).  Other groups have departed either from the canonical MRN power-law size distributions (e.g. lognormals) in order to explain various features of the Galactic extinction curve \citep[e.g.][]{Kim1994,Li2001}, or employed non purely graphite-silicate dust compositions \citep[motivated in large part by observed emission from polycyclic aromatic hydrocarbons; ][]{Jones1990,Siebenmorgen1992,Dwek1997,Li1997,Weingartner2001,Zubko2004,Galliano2011,Jones2017}.

The second major class of theoretical models aims to directly simulate the evolution of dust grains in ISM or galaxy evolution simulations.  These simulations build off of the significant literature modeling single-sized dust in galaxies \citep[e.g.][]{McKinnon2016,Zhukovska2016,McKinnon2017,Popping2017,Vogelsberger2019,Li2019,Dave2019,Vijayan2019} and generally include models for the formation of dust in the ejecta of evolved stars, a range of growth processes, and the destruction of dust.  \citet{Asano2013} and \citet{nozawa2015} developed simplified galaxy one-zone models in which they aggregated the equations that govern dust formation, growth by accretion and coagulation, and destruction by thermal sputtering and shattering in order to develop early numerical models for galaxy grain size distributions.  These, coupled with an assumption for the optical properties of the grains \citep[e.g.][]{Draine1984} result in an extinction curve.  \citet{Hirashita2015} developed a two-size grain model (small and large grains) which has since been applied (or similar variants of this model) to bona fide galaxy evolution simulations by a number of groups  \citep{Gjergo2018,Aoyama2018,Hou2019}.  To date, the only cosmological simulations to include dust with multiple sizes have employed this two-size approximation.  At the same time, a number of studies have implemented dust with multiple ($>2$) grain sizes in idealized, non-cosmological galaxy simulations \citep{McKinnon2018,Aoyama2020}.  While these simulations represent a substantial step forward in their ability to self-consistently evolve the grain size distribution along with the fluid quantities in hydrodynamic galaxy simulations, they are unable to model the cosmological evolution of galaxies, and the attendant physical processes that may impact the dust masses and grain sizes.

In this paper, we develop the first cosmological hydrodynamic galaxy formation model to self-consistently model the evolution of dust grain sizes and masses in the ISM of galaxies over cosmic time.   Here, we employ these simulations to understand, in specific, the origin of the inferred dust grain size distribution and observed extinction curve in Milky Way-mass galaxies at $z \sim 0$.  Our goal is to develop a physical model that understands the observed range in extinction law slopes and bump strengths within the Milky Way and similar galaxies.  This paper is organized as follows. In \S\ref{sec:method}, we summarize the set up of the cosmological simulation, with a particular focus on the model for grain size evolution.  In \S~\ref{sec:results}, we show our major results, and discuss both the origin of the Milky Way extinction law, as well as how and why the bump strengths and slopes vary. We provide discussion in \S~\ref{sec:discuss}, where we compare to other models, as well as  discuss our uncertainties.  In \S~\ref{sec:conclude}, we provide summary.

\section{Methods}
\label{sec:method}

\subsection{Cosmological Galaxy Formation Simulations}

In this paper we have run a series of cosmological simulations with the \simba \ galaxy formation physics suite, though with some significant updates to the dust model that we outline in \S\ref{section:dust_model}. We refer the reader to \citet{Dave2019} for full details, and we summarize the salient points here.  For reviews of broader numerical galaxy formation techniques, see \citet{Somerville2015} and \citet{Vogelsberger2020}.

The primary simulation we use here has $256^3$ dark matter particles and $256^3$ gas elements in a cube of $12\hmpc$ side length, and is run from $z=99$ down to $z=0$.  Because of the small box size, we run $8$ such simulations, varying the random seeds for the initial conditions in order to produce different final galaxy populations.  This enables us to build larger samples of galaxies while maintaining reasonable mass resolution.
We assume a Planck16 \citep{Planck2016} concordant cosmology of $\Omega_m=0.3$, $\Omega_\Lambda=0.7$, $\Omega_b=0.048$, $H_0=68\kmsmpc$, $\sigma_8=0.82$, and $n_s=0.97$. Our run has a minimum gravitational softening length $\epsilon_{\rm min} = 0.25 \hkpc$, mass resolution $1.2\times 10^7 \msolar$ for dark matter particles and $2.28\times 10^6 \msolar$ for gas elements. The system is evolved using a forked version of the \gizmo\ cosmological gravity plus hydrodynamics solver~\citep{Hopkins2015}, in its Meshless Finite Mass (MFM) version. This code, modified from \gad~\citep{Springel2005}, evolves dark matter and gas elements together including gravity and pressure forces, handling shocks via a Riemann solver with no artificial viscosity.

Radiative cooling and photoionisation heating are modeled using the {\sc Grackle-3.1} library~\citep{Smith2017}, including metal cooling and non-equilibrium evolution of primordial elements. Star formation occurs in H$_2$ molecular gas, where the H$_2$ fraction is computed based on the sub-grid model of \citet{Krumholz2009} based on the metallicity and local column density, with minor modifications as described in \citet{Dave2016} to account for variations in numerical resolution.  The star formation rate is given by the H$_2$ density divided by the dynamical time: SFR$=\epsilon_*\rho_{\rm H2}/t_{\rm dyn}$, where we use $\epsilon_*=0.02$~\citep{Kennicutt1998}. These stars drive winds in the interstellar medium. This form of feedback is modeled as a two-phase decoupled wind,  with $30\%$ of wind particles ejected hot, i.e. with a temperature set by the supernova energy minus the wind kinetic energy. The modeled winds have an ejection probability that scales with the the galaxy circular velocity and stellar mass (calculated on the fly via fast friends-of-friends galaxy identification). The nature of these scaling relations follow the results from higher-resolution studies in the Feedback In Realistic Environments zoom simulation campaign \citep[e.g.][]{Muratov2015,Angles2017b,Hopkins2014,Hopkins2018}.

The chemical enrichment model tracks eleven elements (H, He, C, N, O, Ne, Mg, Si, S, Ca, Fe) during the simulation, with enrichment tracked from Type II supernovae (SNe), Type Ia SNe, and Asymptotic Giant Branch (AGB) stars. The yield tables employed are: \citet{Nomoto2006} for SNII yields, \citet{Iwamoto1999} for SNIa yields, and AGB star enrichment following \citet{Oppenheimer2006}. Type Ia SNe and AGB wind heating are also included, along with ISM pressurisation at a minimum level as required to resolve the Jeans mass in star-forming gas as described in \citet{Dave2016}.

\simba\ incorporates black hole physics. Black holes are seeded and grown during the simulation via two-mode accretion. The first mode closely follows the torque-limited accretion model presented in \citet{Angles2017a}, and the second mode uses Bondi accretion, but solely from the hot gas component. The accretion energy is used to drive feedback that serves to quench galaxies, including a kinetic subgrid model for black hole feedback, along with X-ray energy feedback. \simba\ additionally includes a dust physics module to track the life-cycle of cosmic dust; we next describe this model, as well as improvements that we make to enable this study.

\subsection{Dust Model}
\label{section:dust_model}
In the original \simba\ model, we model dust approximated as a single grain size that passively advects with the gas \citep{Li2019}.  In this paper, we have significantly updated this model to instead treat dust as its own particle that experiences grain-gas drag and gravity. Each particle contains a collection of grains with a fixed mass density $\rho_{\rm gr} = 2.4$~g~cm$^{-3}$ \citep{Draine2003}, and has a distribution of grain sizes that evolves owing to a range of physical processes.

\subsection{Dust Production}
A fraction of metals returned to the ISM by the ejecta of AGB stars and SNe II may condense into dust. We neglect 
the condensation of SN Ia as it is a negligible source (see e.g. \citealt{Nozawa2006,Dwek2016,Gioannini2017}). To model the condensation, we follow the prescription of \citet{Dwek1998} with updated 
condensation efficiencies. We refer readers to \citet{Li2019} for details.

Dust particles are stochastically created assuming that the creation is a Poisson process, following \citet{McKinnon2018}. During a time-step, the probability of a star or gas particle of mass $M$ spawning a dust 
particle with mass $M_d=\beta M$ ($\beta=0.02$ in our simulation) is
\begin{equation}
p_d = \frac{M}{M_\dust}[1-\exp (- \frac{m^j_\dust}{M})]
\label{eq:1}
\end{equation}
where $m_\dust^j$ is the expected dust mass produced by the $j$th stellar process (SNII or AGB stars). We assume that the total carbon mass corresponds to the graphite mass, and the remainder goes to silicates, A random number is drawn between 0 to 1. If the number is smaller than $p_d$ then a dust particle is created.

Once the decision has been made to create a dust particle, the initial grain size distribution is assigned according 
to the type of the stellar process. To represent the distribution, we divide [$\log a_{\rm min}$,$\log a_{\rm min}$] 
into $N_{\rm bin} = 41$ equally sized bins, where $a_{\rm min}$~=~$10^{-4}$~$\micron$ and $a_{\rm max}$~=~1~$\micron$. Throughout the work, we apply piece-wise constant discretisations to the grain size distributions. 

We assume the initial grain size distribution is
\begin{equation}
    \frac{\partial n}{\partial a} = \frac{C}{a^p} \exp \left( \frac{\ln^2(a/a_{0})}{2 \sigma ^2} \right),
\end{equation}
where C is a normalization constant, $a_0 = 0.1 \micron$. $(p,\sigma)=(4,0.47)$ for dust produced by AGB stars and $(p,\sigma)=(0,0.6)$ for SNII, following the work by  \citet{Asano2013b}.

 We then 
draw $N$~=~$10^4$ random numbers between $\log a_{\rm min}$ and $\log a_{\rm max}$ using the Metropolis-Hasting algorithm to sample the initial size distribution, with $N_k$ numbers in the $k$th bin. Note that we assume that the silicate grains and graphite grains have the same size distributions in this work, and the distributions are not evolved separately for different species.

\subsection{Dust Growth}
Once dust particles are produced, they are able to grow by accreting neighboring gas-phase metals. Following 
{Dwek 1998}, the growth rate of grain radius $a$ can be expressed as:
\begin{equation}
\left( \frac{\diff a}{\diff t} \right)_{\rm grow}=\frac{a}{\tau_{\rm accr}},
\label{eq:2}
\end{equation}
where $\tau_{\rm accr}$ is the the characteristic accretion timescale. Following 
\citet{Hirashita2000}, which assumes the process is a two-body collision, the timescale $\tau_{\rm accr}$ is 
\begin{equation}
\tau_{\rm accr} = \tau_{\rm ref} \left(\frac{a}{a_{\rm ref}}\right) \left( \frac{\rho_{\rm ref}}{\rho_\gas} \right) {\left(\frac{T_{\rm ref}}{T_\gas} \right)} ^{\frac{1}{2}} \left( \frac{Z_\odot}{Z_\gas} \right).
\label{eq:3}
\end{equation}
where $a_{\rm ref} = 0.1 \micron$ and $\rho_\gas$, $T_\gas$ and $Z_\gas$ are the neighboring gas density, temperature and metallicity, respectively. 
$\tau_{\rm ref}$, $\rho_{\rm ref}$, $T_{\rm ref}$ and $Z_{\rm ref}$ are the reference values correspondingly, which have $(\tau_{\rm ref}/{\rm Gyr}, \rho_{\rm ref}/\mH \cm^{-3}, T_{\rm ref}/{\rm K}, Z_{\rm ref}) = (0.03,100,20,0.0134)$.

To compute the grain growth time scale in Equation~(\ref{eq:3}), $\rho_\gas$, $T_\gas$ and $Z_\gas$ are evaluated by smoothing properties of $N_{\rm ngb} \sim 32$ neighboring gas particles in a kernel-weighted way. To this end, we first iteratively solve the smoothing length $h_{\rm dg}$ via 
\begin{equation}
N_{\rm ngb} = \frac{4\pi h^3_{\rm dg}}{3}\sum_{r_i < h_{\rm dg}} W(r_i,h_{\rm dg}),
\label{eq:5}
\end{equation}
where $r_i$ is the distance from the dust particle to the $i$th gas particle within a sphere with a radius $h_{\rm dg}$ and $W(r,h)$ is the cubic spline kernel.

Then we can compute gas properties via
\begin{equation}
\rho_\gas = \sum_{i=1}^{N_{\rm ngb}} M_i W(r_i,h_{\rm dg})
\label{eq:6}
\end{equation}
and
\begin{equation}
T_\gas = \frac{\sum_{i=1}^{N_{\rm ngb}} M_i T_i W(r_i,h_{\rm dg})}{\rho_\gas}.
\label{eq:7}
\end{equation}
Other gas properties needed for dust physics are evaluated in the same manner as Equation~(\ref{eq:7}).

\subsection{Dust Destruction}
\subsubsection{Thermal Sputtering}
Dust grains can be eroded by colliding with thermally exited gas especially in hot halos, a process known as "sputtering" \citep{Barlow1978, Draine1979a, Tielens1994}. In this work, we adopt an analytic approximation derived by \citet{Tsai1995}:
\begin{equation}
\left( \frac{\diff a}{\diff t} \right)_{\rm sp} = -\frac{a}{\tau_{\rm sp}},
\label{eq:8}
\end{equation}
where the characteristic time scale
\begin{equation}
\tau_{\rm sp}  \sim (0.17\ {\rm Gyr}) \left( \frac{a}{a_{\rm ref}} \right) \left( \frac{10^{-27}{\rm \ g\ cm^{-3}}}{\rho_g} \right) \left[ \left( \frac{T_0}{T} \right)^{\omega}+1 \right],
\label{eq:9}
\end{equation} 
where $\omega$~=~$2.5$ controls the low-temperature scaling of the sputtering rate and 
$T_0\ =\ 2 \times 10^6$~K is the temperature above which the sputtering rate flattens. 

\subsubsection{Dust Destruction via SN Shocks}
In addition to thermal sputtering, SN blast waves offer another approach to destroying dust grains by enhancing inertia and thermal sputtering \citep{Dwek1980,Seab1983,McKee1987,McKee1989}.  The SN shocks shifts the grain size distribution to smaller sizes. We follow \citep{McKinnon2018}, who build off of \citet{Yamasawa2011} and \citet{Asano2013} to determine the evolution of grain size distribution caused by SN shocks. This method is parameterized by a conversion efficiency $\xi(a,a')$ such that $\xi(a,a') \diff a$ denotes the fraction  of grains starting with sizes $[a',a'+da]$ that end up with sizes $[a,a+da]$. The $\xi(a,a')$ values are calculated using a detailed model of dust destruction in SN blast waves developed by \citet{Nozawa2006}. The rate of change of number of grains in the $k$th bin is
\begin{equation}
\left( \frac{\diff N_k}{\diff t} \right)_{\rm de} = \frac{\gamma M_s}{M_g} \left( \sum_{i=1}^{N_{\rm bin}} N_i(t) \xi(a_k,a_i) \diff a - N_k(t) \right), 
\label{eq:11}
\end{equation}
where $M_g$ is the neighboring gas mass, $\gamma$ is the neighboring SN II rate, and $M_s$ is the mass of neighboring gas shocked to at least 100 km/s per SN event. Because our simulations do not resolve the multiphase ISM, we apply the Sedov-Taylor solution to a homogeneous medium of $\nH=0.13$~cm$^{-3}$ (the minimum SF threshold density of our simulations) following \citet{McKee1989}, yielding:
\begin{equation}
M_s = 6800 E_{\rm SNII,51}\left(\frac{v_s}{100\ {\rm km\ s^{-1}}}\right)^{-2} M_\odot,
\label{eq:12}
\end{equation}
where $E_{\rm SNII,51}$ is the energy released by a SN II in units of 10$^51$~erg, and $v_s\sim 100$~km~$s^{-1}$ is the shock wave speed. The resulting change rate of mass of grains in bin $k$ is
\begin{equation}
\left( \frac{\diff M_k}{\diff t} \right)_{\rm de} =  \frac{\gamma M_s}{M_g} \left\lbrace \sum_{i=1}^{N_{\rm bin}} \left[ N_i(t) \xi(a_k,a_i) \left( \frac{\pi \rho_{\rm gr} a^4}{3} \right) \right]_{a_k}^{a_{k+1}} - M_k(t) \right\rbrace
\label{eq:13}
\end{equation}

\subsubsection{Grain Shattering and Coagulation}
Grain-grain collisional processes including shattering and coagulation, though conserve the dust mass, could significantly shape the grain size distributions. In this work, we follow the approach of \citet{McKinnon2018} and the mass evolution for grain size bin $k$ is

\begin{equation}
\begin{split}
\frac{\diff M_k}{\diff t} = &- \frac{\pi \rho_\dust}{M_\dust}
(\sum_{k=0}^{N-1} v_{\rm rel} (a_i,a_k) \mathbb{1}_{v_{\rm rel} > v_{\rm th}} (a_i,a_k) m_i I^{i,k} \\ 
&- \frac{1}{2} \sum_{k=0}^{N-1} \sum_{j=0}^{N-1} v_{\rm rel} (a_k,a_j) \mathbb{1}_{v_{\rm rel} > v_{\rm th}} (a_k,a_j) m^{k,j}_{\rm col}(i) I^{k,j} ),
\label{eq:14}
\end{split}
\end{equation}
where $\rho_\dust$ is the mass density of dust, $v_{\rm rel}(a_i,a_k)$ is the relative velocity of two grains at grain size bins $i$ and $k$ respectively (we assume grains with in one dust particle and one size bin have the same velocity for simplicity), $m_i$ is the average mass of a grain in bin $i$, $m^{k,j}_{\rm col}(i))$ is the resulting mass entering bin $i$ due to the collision between grains in bins $k$ and $j$, $v_{\rm th}$ is the threshold velocity where shattering or coagulation can happen, and 
\begin{equation}
\begin{split}
I^{k,j} = \int_{a_k}^{a_{k+1}} \int_{a_j}^{a_{j+1}} \frac{N_k N_j}{(a_{k+1}-a_k)(a_{j+1}-a_j)}(a_1 + a_2)^2 \diff{a_2}\diff{a_1}.
    \label{eq:15}
\end{split}
\end{equation}

To calculate the relative velocity $v_{\rm rel}$, we follow the calculation by \citet{Hirashita2019} where grain velocity is set by the drag force of the turbulent gas flow, but assume the turbulent velocity has a supersonic power spectrum instead of the Kolmogorov power spectrum, giving the velocity dispersion of grains with radii $a$:

\begin{equation}
\begin{split}
\sigma_{\rm gr}  = 0.06 \left(\frac{v_J}{c_\gas}\right) ^2 \left(\frac{a}{0.1 \micron} \right) \left(\frac{\rho_\gas}{1 \cm^{-3} \times \mH} \right)^{-\frac{1}{2}} \left(\frac{\rho_{\rm gr}}{2.4 {\rm g\ \cm^{-3}}} \right) \ {\rm km\ s^{-1}},
\end{split}
\end{equation}
where $c_\gas$ is the local speed of sound and $v_J=0.7 (L_J/1{\rm pc})^{1/2} \kms$ \citep{Solomon1987} is the turbulent velocity at the size of eddies with the Jeans Length $L_J \equiv (\pi c_\gas^2/G\rho_\gas)^{1/2}/2$.
The relative velocity of grains are then calculated via $x$, $y$, $z$ components of grain velocities randomly drawn from Gaussian distributions $N(0,\sigma_{\rm gr}^2/3)$.

For shattering,  \citet{Jones1996} uses a threshold velocity $v_{\rm th} = 2.7\kms$ for silicate grains and $v_{\rm th} =1.2\kms$ for graphite grains. We adopt $v_{\rm th} =2\kms$ for all grain species, for we do not track detailed evolution of multiple grain species. The computation of $m^{k,j}_{\rm col}(i)$ follows Section~2.3 of \citet{Hirashita2009}, which considers partial or complete fragmentation of colliding grains.

For coagulation, the threshold velocity $v_{\rm th}$ is computed via Equation~8 of \citet{Hirashita2009}, dependent of grain sizes $a_k$ and $a_j$ of two colliding grains, and
\begin{equation}
    m^{k,j}_{\rm col}(i) =
    \begin{cases}
    m_k + m_j, &\log a_i + \frac{1}{2} \Delta (i) \leq \frac{m_k+m_j}{4\pi \rho_{\rm gr}/3} < \log a_{i+1} - \frac{1}{2} \Delta (i),\\
    0, & \text{otherwise},
    \end{cases}
    \label{eq:16}
\end{equation}
where $\Delta(i)\equiv\log a_{i+1} - \log a_{i}$.

\subsubsection{Dust Consumption via Star Formation}
The mass of dust particles is reduced when star particles are created in the neighborhood, a process known as "astration". To evaluate the amount of dust mass consumed by star particles, we first compute the weight for the $i$th neighboring gas or star particle within the sphere with a radius $h_{\rm dg}$:
\begin{equation}
    w_i = m_i W(r_i,h_{\rm dg})
\label{eq:17}
\end{equation}
where $h_{\rm dg}$ is determined by Equation~(\ref{eq:5}). Then a fraction $f_j$ of mass of the dust particle is consumed by the $j$th nascent star particle, where
\begin{equation}
    f_j = \frac{w_j}{\sum_{r_i < h_{\rm dg}}w_i}.
\label{eq:18}
\end{equation}
The metal mass and momentum are assumed to be conserved during this process.

\subsection{Dust Dynamics}
The motion
of dust particles follow the pressureless fluid dynamics, interacting with the gas fluid via gravity and a drag force given by:

\begin{equation}
\frac{\diff v_g}{\diff t} = a_{\rm drag} + a_{\rm ex},
\end{equation}
where $a_{\rm ex}$ denotes external sources of acceleration (in particular gravity for our simulations), and v$_{g}$ is the gas velocity. The acceleration caused by the drag force $a_{\rm drag}$ is given by
\begin{equation}
a_{\rm drag}  = -\frac{v_{\rm d} - v_{\rm g}}{t_s},
\end{equation}
where $v_{d}$ is the dust velocity, and the stopping time $t_s$ is given by
\begin{equation}
t_s  = \frac{M_{\rm d} \rho_{\rm g}}{K_s (\rho_\gas + \rho_\dust)} \sim \frac{M_{\rm d}}{K_s}
\end{equation}
as the dust density $\rho_\dust$ typically satisfies $\rho_\dust / \rho_\gas \ll 1$ in the cosmological simulations.  Here $K_s$ is the drag coefficient (described below).

In the cosmological simulations, the typical radii of dust grains $a \ll 9\lambda / 4$ where $\lambda$ is the mean free path of gas particles, corresponding to the Epstein regime \citep{Epstein1924}. The drag coefficient in this regime is given by
\begin{equation}
K_s = \frac{8\sqrt{2\pi} c_\gas a^2 \rho_\gas}{3\sqrt{\gamma}}
\end{equation}
where $a$ is the grain radius, and $\gamma$ is the adiabatic index. This gives
\begin{equation}
t_s \sim \frac{M_{\rm d}}{K_s} = \frac{\sqrt{\pi \gamma} a \rho_{\rm gr}}{2\sqrt{2} \rho_\gas c_s}.
\label{eq:ts}
\end{equation}
This assumes subsonic dust-to-gas relative velocities and should be corrected by the following fit \citep{Draine1979a}:
\begin{equation}
t_s =  \frac{\sqrt{\pi \gamma} a \rho_{\rm grain}}{2\sqrt{2} \rho_\gas c_s} \left( 1 + \frac{9\pi}{128} \left| \frac{v_\dust - v_\gas}{c_s} \right| ^2  \right)^{-\frac{1}{2}}
\end{equation}
in order to be applied to supersonic dust-to-gas relative velocity.

The time integration follows the semi-implicit time-stepping approaches detailed in \citet{Hopkins2016} 
to lift the strict time-stepping requirement $\diff t < t_s$ for an explicit integrator when the stopping time $t_s$ is much smaller than the time-scale of other accelerations, which is typical of the cosmological simulations where dust couples with gas in most regions. The integrator can be expressed by
\begin{equation}
v_\dust (t + \diff t) = \tilde{v}_\dust (t+\diff t) - \xi [ \tilde{v}_\dust (t+\diff t) - \tilde{v}_\gas (t+\diff t)] + [\xi(\diff t + t_s) - \diff t] + \frac{\nabla P}{\rho_\gas},
\end{equation}
where $\tilde{v}$ denotes the velocity at time $t+\diff t$ after non-drag kicks are applied but before the drag force is applied, and $\xi = 1 - \exp({-\diff t / t_s})$. This gives a Courant-Friedrichs-Lewy (CFL) type time-step
\begin{equation}
\diff t_{\rm CFL} = \frac{C_{\rm CFL} h_\dust}{\sqrt{c_s^2 + |v_\dust - v_\gas|^2}},
\end{equation}
where $h_\dust$ is the smoothing length for dust particles.

In order to track the formation and evolution of dust grains and their sizes, we implement dust production via 
condensation of stellar ejecta, dust growth via accretion of gas-phase metals, dust destruction via shock waves 
and thermal sputtering, and grain coagulation and shattering. The choice of free parameters is shown in Table~\ref{tab:param}, with further discussion in \citet{Li2019}.


\begin{table*}
	\centering
	\caption{Simulation Free Parameters\label{tab:param}}
	\begin{tabular}{lcc} 
		\hline
		Parameter & Description & Value\\ 
		\hline
		Thermal sputtering & &  \\
		$\rho_{\rm gr}$ & Density of solid matters within grains (g cm$^{-3}$)$^a$ & 2.4  \\
		Production & &  \\
		$\delta^{\rm AGB,C/O>1}_{i,\rm dust}$ & Condensation efficiency$^b$ & 0.2 for $i$ = C \\ 
		                                     &                        &  0 otherwise        \\
		$\delta^{\rm AGB,C/O<1}_{i,\rm dust}$ &                       & 0 for $i$ = O  \\
		                                      &                       &  0.2 otherwise\\ 
		$\delta^{\rm SNII}_{i,\rm dust}$     &                         & 0.15 for $i$ = C\\  
		                                      &                       &  0.15 otherwise\\  
		Growth & & \\
		$\rho^{\rm ref}$ & Reference density (g cm$^{-3}$) & $2.3\times 10^{-22}$ \\
		$T^{\rm ref}$ & Reference temperature (K) & 20  \\
		$\tau_{\rm g}^{\rm ref}$ & Growth time-scale with $T=T^{\rm ref}$ and $\rho = \rho^{\rm ref}$ (Myr)$^c$   & 10\\ 
		Destruction (SN Shocks) & &\\
		$E_{\rm SN,51}$ & Energy per SN ($10^{51}$ erg)$^d$ & 1.0\\
		\hline
	\end{tabular}
	\\$^a$ \citet{Jones1996, Draine2003}.
	\\$^b$ \citet{Dwek1998,McKinnon2017,Popping2017}
    \\$^c$ \citet{Dwek1998,Zhukovska2014,McKinnon2017,Popping2017}
    \\$^d$ \citet{McKee1989}.
\end{table*}

\subsection{Galaxy Identification and Tracking}
Halos are identified on the using a $3$-D friends-of-friends algorithm within \gizmo, with a linking length of $0.2$ times the mean inter-particle spacing.  Galaxies are identified via a $6$-D friends-of-friends technique within the publicly available galaxy analysis tool {\sc caesar}\footnote{\url:https://github.com/dnarayanan/caesar}. The minimum number of baryonic particles for an identified galaxy is 24, leading to a minimum baryonic mass $\sim 5.47 \times 10^7 \msolar$. We consider galaxies at $z=0$ within $0.3$ dex of a stellar mass of $6 \times 10^{10}$ M$_\odot$ \citep{licquia2015} and a halo mass of $1.6 \times 10^{12}$ M$_\odot$ \citep{boylankolchin13a} as reasonable analogs to the Milky Way.  In total, we identify 12 galaxies at $z=0$ within this mass range. Furthermore, we trace back their most massive progenitors at each earlier snapshot to track the evolutionary history of these galaxies.


%
%
%
%

\subsection{Validation: Global dust-to-gas ratios}
\label{sec:benchmark}
\begin{figure}
\includegraphics[width=0.5 \textwidth]{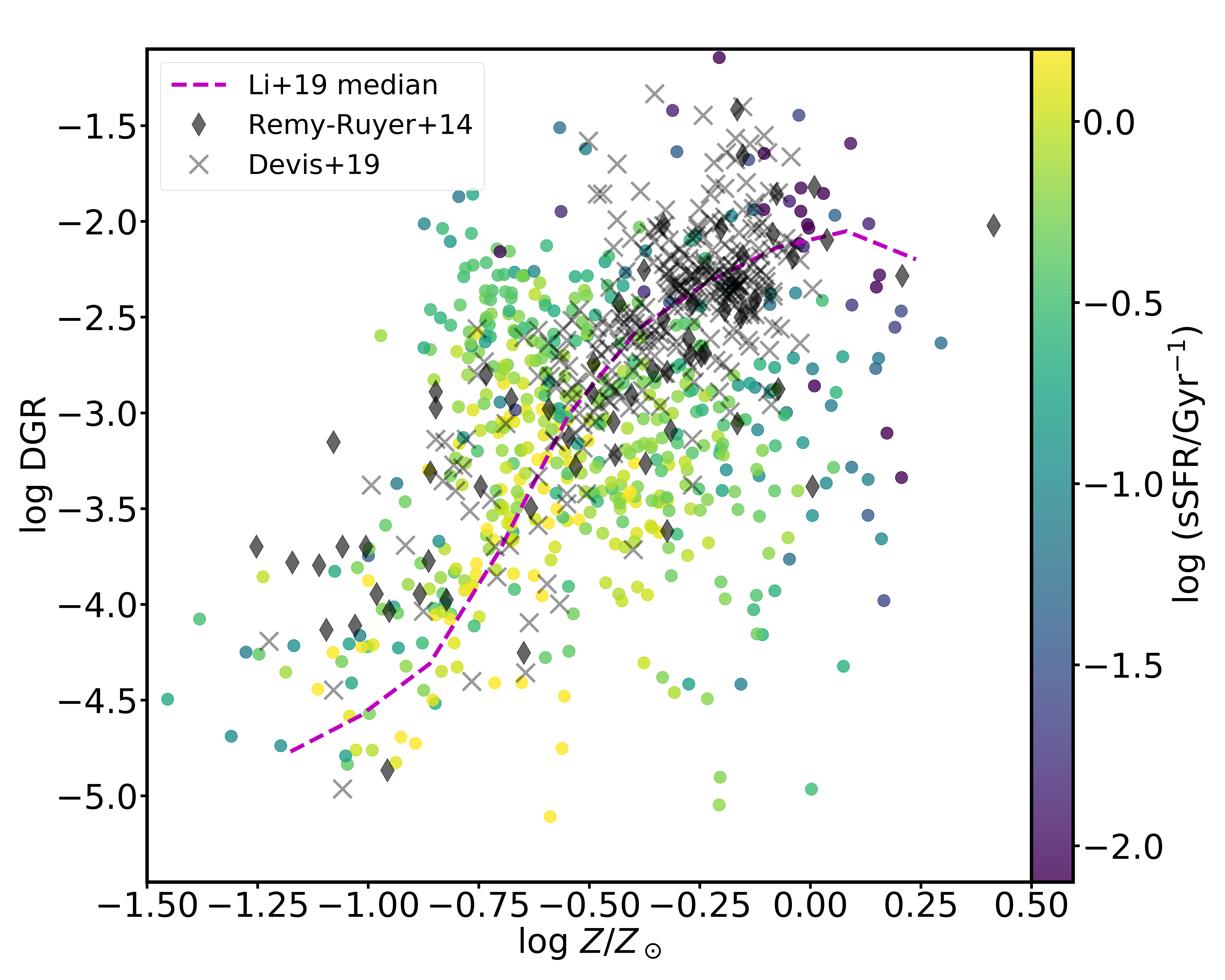}
\caption{Verification of our methodology and model results by comparing the simulated dust-to-gas ratio vs metallicity relation to observed galaxies at $z=0$.  The yellow$\rightarrow$purple data (viridis color map) show all of our simulated galaxies at $z=0$ (i.e. not just Milky Way-mass), while the black diamonds and crosses show the observational constraints by \citet{Remy-Ruyer2014} (assuming a metallicity-dependent CO-to-H$_{2}$ conversion factor) and \citet{DeVis2019}, respectively. The colors of the simulated data quantify their specific star formation rate.  The dashed magenta line denotes the running median of the dust-to-gas ratio versus gas-phase metallicity relation from our previous passive dust model \citep{Li2019}. Our model reproduces the general observed trend of increasing dust-to-gas ratio and metallicity. 
\label{figure:benchmark}}
\end{figure}

Before proceeding, we present a validation of this model as applied to our simulations. The scaling relation between dust-to-gas ratio (DGR) and gas-phase metallicity represents an important constraint on models of the dust life-cycle.   In Figure~\ref{figure:benchmark}, we plot the modeled $z=0$ dust to gas mass ratio against the galaxy gas phase metallicity $Z$ from our simulation.  The colored points (viridis map) show our model galaxies at $z=0$, while the  black diamonds and grey crosses show the observational constraints from \citet{Remy-Ruyer2014} (assuming a metallicity-dependent CO-to-H$_{2}$ conversion factor) and \citet{DeVis2019}.  We see excellent correspondence between our model DGR and the observational constraints at both high and low metallicity. 

We see roughly a linear increase of DGR as a function of $Z$ at $Z \lesssim 0.15 Z_\odot$ and $Z \gtrsim 0.5 Z_\odot$ which correspond to the regimes dominated by dust production and dust growth, respectively. There is a nonlinear rise from $Z \sim 0.15 Z_\odot$ to $Z \sim 0.5 Z_\odot$ which corresponds to the transition from production-dominated regime to growth-dominated regime. 
We also compare this scaling relation to the relation we get from the passive dust model of \citet{Li2019}, the running median of which is denoted by dashed magenta lines. The relation from current model generally follows the trend we get from the passive dust model. We see that there is large scatter as dust growth becomes dominant. This indicates the impact of the variance in mass-averaged grain sizes for dust particles instead of a single grain size $a=0.1 \micron$ assumed by our previous work (c.f. Equation~\ref{eq:3}).


\section{Results}
\label{sec:results}
\subsection{Extinction Curves in Milky Way Mass Galaxies}
\label{sec:dsf}
\begin{figure}
\includegraphics[width = 0.5 \textwidth]{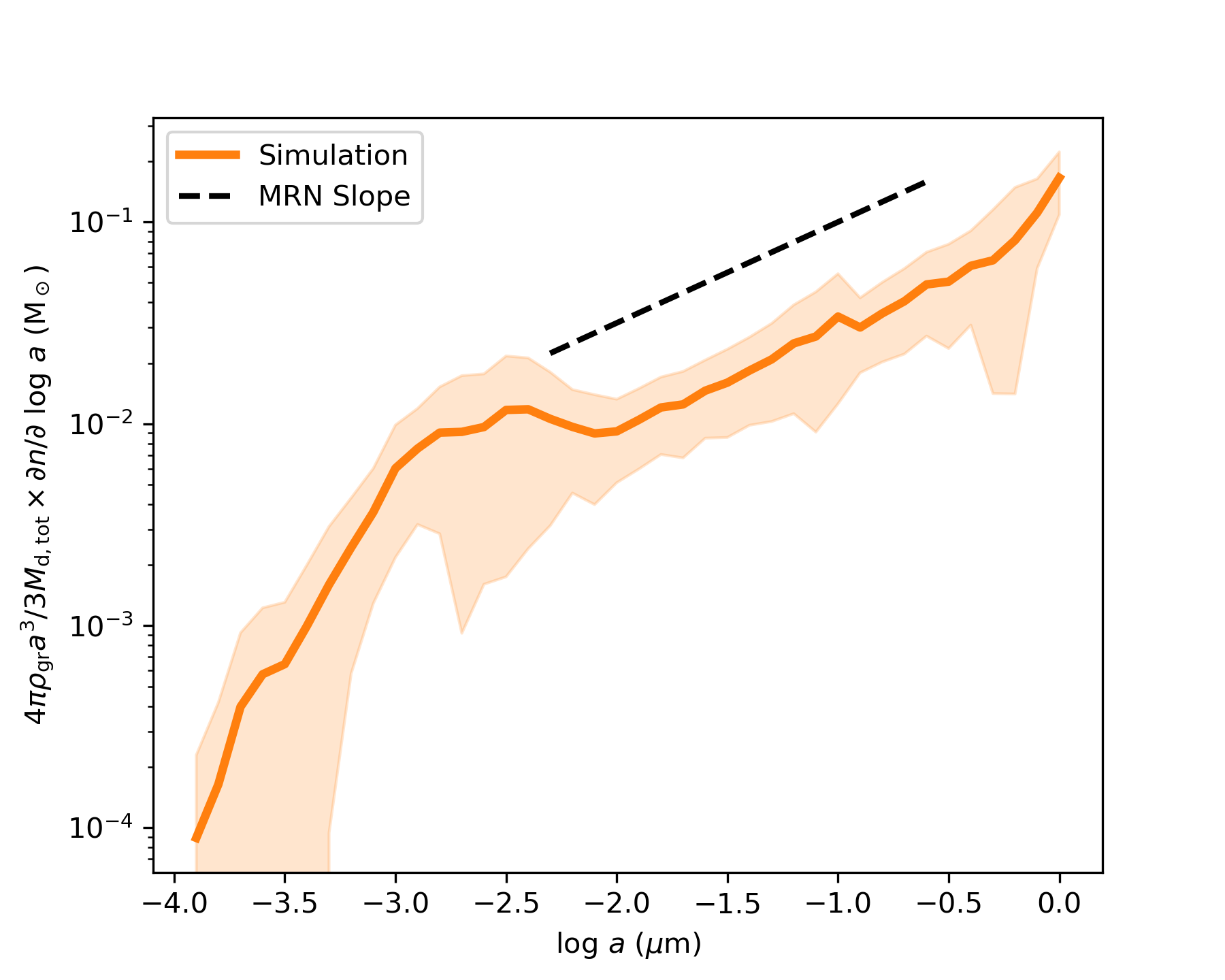}
\caption{Mass-weighted grain size distributions for Milky Way analogs in our simulation  (the peach-colored line shows the mean, while the shaded region shows the $1\sigma$ dispersion amongst our sample of galaxies).  Our model Milky Way analogs at $z=0$  produce a diverse range of grain size distributions, though with slopes comparable to a traditional MRN distribution (dashed line -- note, the MRN distribution normalization is arbitrary, and we manually offset it from our model galaxies to enhance clarity).  This said, the diversity in the small grain size distributions drives variation in the UV/optical slopes of the extinction curves, while the bump strengths are more closely tied to the fraction of graphites vs silicates in a galaxy.}
\label{fig:dmsf}
\end{figure}

We first ask the question: do our model galaxies with a comparable stellar mass and halo mass as the Milky Way have extinction curves comparable to observed constraints?  In Figure~\ref{fig:dmsf}, we plot the dust size distribution for all of our Milky Way analogs at $z=0$, and compare this to a \citet{Mathis1977} ("MRN") powerlaw slope.  While there is significant diversity in the grain size distributions (a topic we will return to later in this paper), the bulk of the distribution functions have slopes comparable to the MRN slope in the size ranges of interest. 
The primary differences in the curves are in the lowest size bins, which, as we will show, have to do with the dust growth history that most closely ties to the metal enrichment history in the galaxy.
We note that our simulations have not been tuned to reproduce this result, but rather the size distributions are a natural consequence of two dominant competing processes i.e. grain growth and destruction processes in our model. We discuss this in more detail shortly. 

In order to compute the extinction curve from the model grain size distributions,  we require knowing the extinction efficiencies (i.e. the ratio of the extinction to geometric cross sections) of our grains.  Here, we assume the models of \cite{Laor1993}, who quantify these cross section ratios for silicates and graphites.  We therefore require assuming a silicate to graphite abundance ratio alongside our computed grain size distributions.  Within our dust super-particles, we assume that the graphite mass corresponds to the total carbon mass, and the remainder is silicates. We further assume that the silicate grains and graphite grains have the same size distribution.  We describe the computation of extinction curves in more detail in Appendix~\ref{appendix:extinction}.  In Figure~\ref{fig:ext}, we show the extinction curve for grains with an MRN size distribution that are comprised of pure graphites (magenta line) and pure silicates (blue line) in order to help the reader interpret our model results.  Note that we use the average grain-size distributions of all dust particles in each galaxies to derive extinction curves instead of particles along a certain line of sight. 

In Figure~\ref{fig:ext}, we show the derived extinction curves from our Milky Way-mass galaxies.  The yellow line and the shaded region shows the mean extinction curve and the standard deviation of the galaxies, while the dashed lines show the range of observationally derived extinction curves as parameterized by $R_{\rm V}$ in the Galaxy \citep{Cardelli1989} and the sightline averaged result from \citet{Fitzpatrick2007}. We additionally show the extinction laws assuming an MRN grain size distribution and a grain composition with 100\% graphites or silicates. This demonstrates that a large graphite to silicate mass ratio contributes to strong $2175$\AA\ bumps while a small ratio leads to a bump-less extinction curve.  Generally, our model Milky Way-mass galaxies show diversity in their extinction curves, though demonstrate excellent correspondence with the observed range in the Milky Way.  We see a range of slopes and bump strengths.  In the remaining subsections, we unpack the origin of this diversity in the curves.

\begin{figure}
\begin{minipage}{0.52\textwidth}
\includegraphics[width = 1 \textwidth]{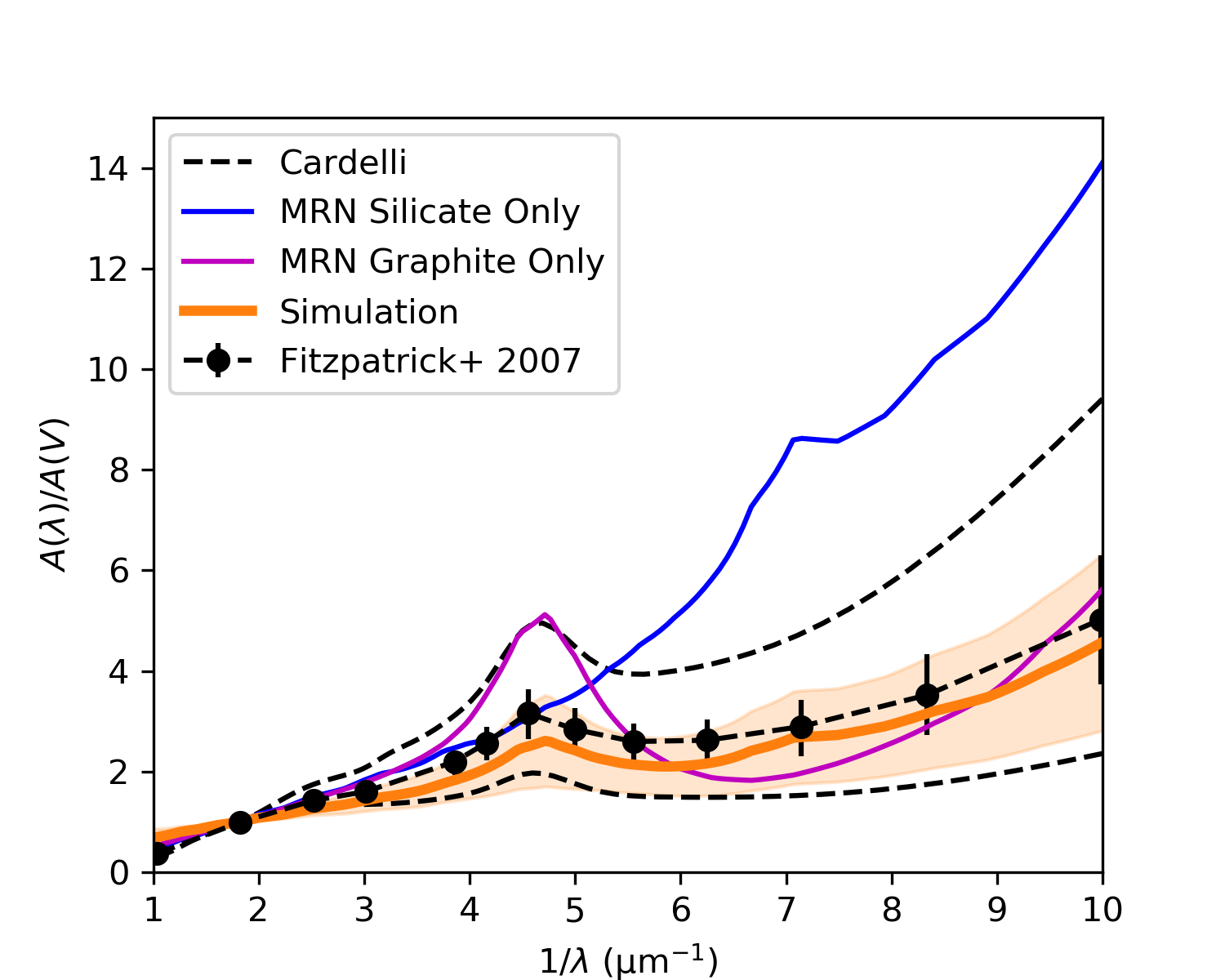}
\end{minipage}
\caption{Model Milky Way mass galaxies formed at $z=0$ in our model (the mean value and the standard deviation denoted by a solid peach-colored line and a shaded region) exhibit dust extinction laws comparable to the range observed in the Milky Way.  The dashed lines denote the bounds of the \citet{Cardelli1989} inferred curves (for a $R_{\rm V}$ range of $[2,5]$), while the dash-dot line denotes the average \citet{Fitzpatrick2007} constraint. The extinction curves assuming an MRN grain size distribution and a grain composition with 100\% graphites or silicates are denoted by magenta and blue lines respectively. These curves are produced by convolving the grain size distributions (c.f. Figure~\ref{fig:dmsf}) with assumed extinction efficiencies of the dust grains for graphites and silicates \citep{Laor1993}.  That the slopes and bump strengths match those of the Milky Way is a reflection of the dominance of grain growth over destruction processes in metal rich environments. See text for details.
\label{fig:ext}}
\end{figure}

\subsection{What Drives the Diversity of Extinction Curves?}
\label{sec:ext}

To characterize the extinction curves and facilitate the comparison amongst different curves quantitatively, we follow the review of \citet{Salim2020} in defining two parameters: the overall UV-optical slope $S$ and the 2175 \AA \  absorption bump strength $B$.  

As a high-level characterization of a curve, the overall UV-optical slope is defined as the ratio of extinction at 1500 \AA \  and in the $V$-band:
\begin{equation}
    {\rm S} \equiv A_{1500} / A_{V},
    \label{eq:slope}
\end{equation}
which generally reflects the relative extinction in the UV band compared to the optical band.  The bump strength is defined as the ratio of extra extinction due to the bump at $2175$ \AA\ to the total extinction at $2175$ \AA\:
\begin{equation}
    B \equiv A_{\rm bump} / A_{2175,0},
    \label{eq:bump}
\end{equation}
where the extinction due to the bump can be estimated by
\begin{equation}
    A_{\rm bump} = A_{2175} - A_{2175,0},
\end{equation}
where the base-line extinction in the absence of the bump can be estimated by
\begin{equation}
A_{\rm 2175,0} \equiv (0.33A_{1500} + 0.67A_{3000}).
\end{equation}
\begin{figure}
\centering
\includegraphics[width=0.48\textwidth]{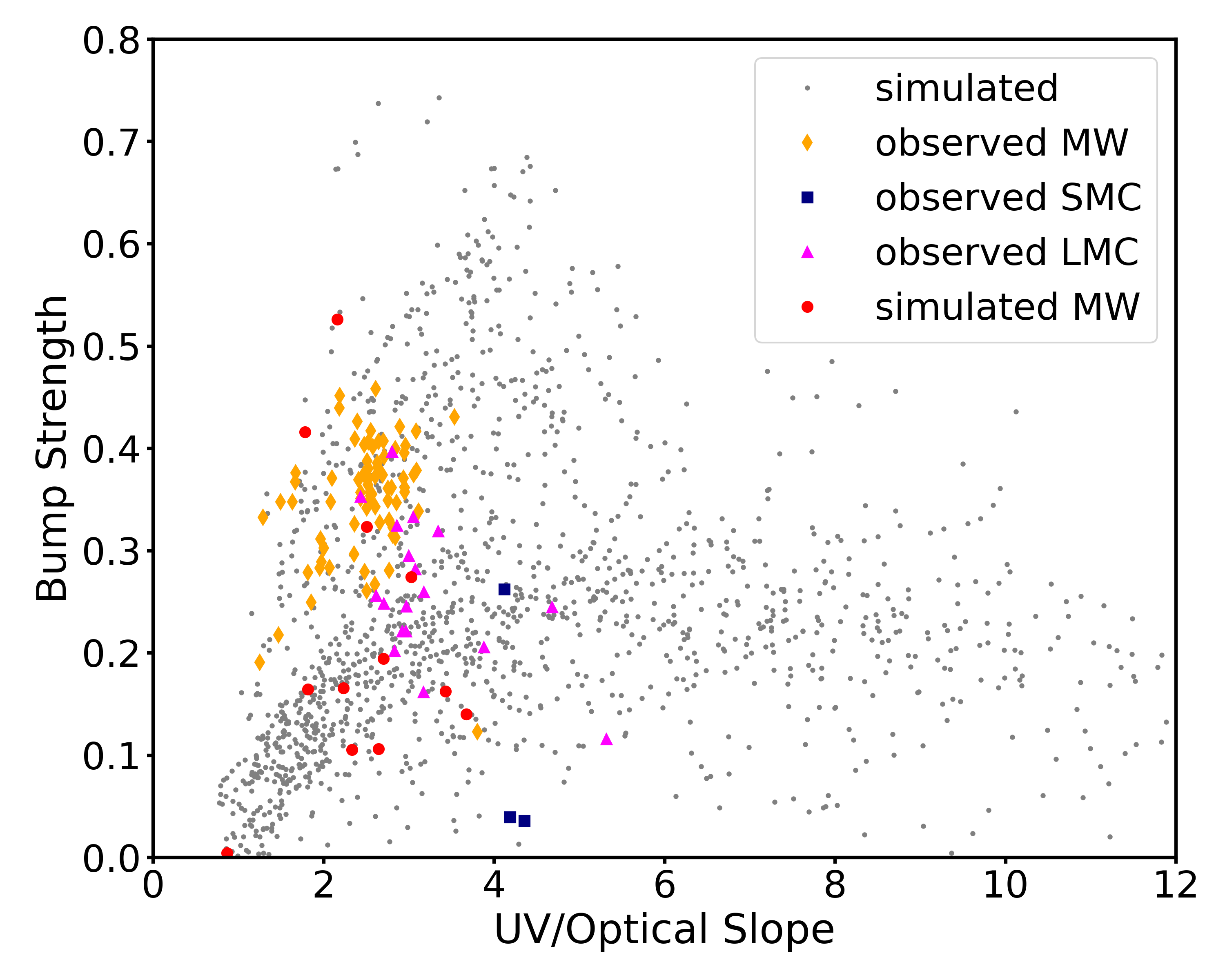}
\caption{ UV bump versus optical slope relation (top) and UV slope vs optical slope) bottom.  Grey points show all galaxies in our model, and orange, blue and pink points show the {\it observed} Milky Way, SMC and LMC data points \citep{Fitzpatrick1990,Fitzpatrick1999,Clayton2000,Fitzpatrick2007, Fitzpatrick2009,Nataf2016,Gordon2003}.  We show all of our galaxies simply to build statistics.  The slopes in our models vary monotonically with metallicity, while the bump strength depends on the graphite to silicate ratio (which shows substantial dispersion at $z=0$).  As a result, in our simulations there is little bump-slope relation in Milky Way analogs.  At the same time, individual observed sightlines in the Galaxy may have a similar graphite to silicate ratio, and therefore, a loose trend between the bump and slope \citep{Salim2020}. Our model cannot account for the small bump strengths of the observed LMC and SMC (see \S~\ref{section:caveats} for details). } 
\label{fig:bump_slope}
\end{figure}
In Figure~\ref{fig:bump_slope}, we show the bump strength versus optical slope relation and UV slope versus optical slope relation for all of our model galaxies (i.e. not just Milky Way-mass galaxies).
We additionally show the observational constraints in both spaces for the Galaxy and Magellanic Clouds by \citet{Fitzpatrick1999, Fitzpatrick2007, Fitzpatrick2009,  Clayton2000, Nataf2016} and \citet{Gordon2003} as light grey points.  Our model Milky Way analogs have a comparable scatter in bump-slope space as the observed Galactic sightlines, though a larger dynamic range (as we will show, this larger dynamic range owes to variations in the graphite to silicate ratio).

\begin{figure}
    \centering
    \includegraphics[width=0.5 \textwidth]{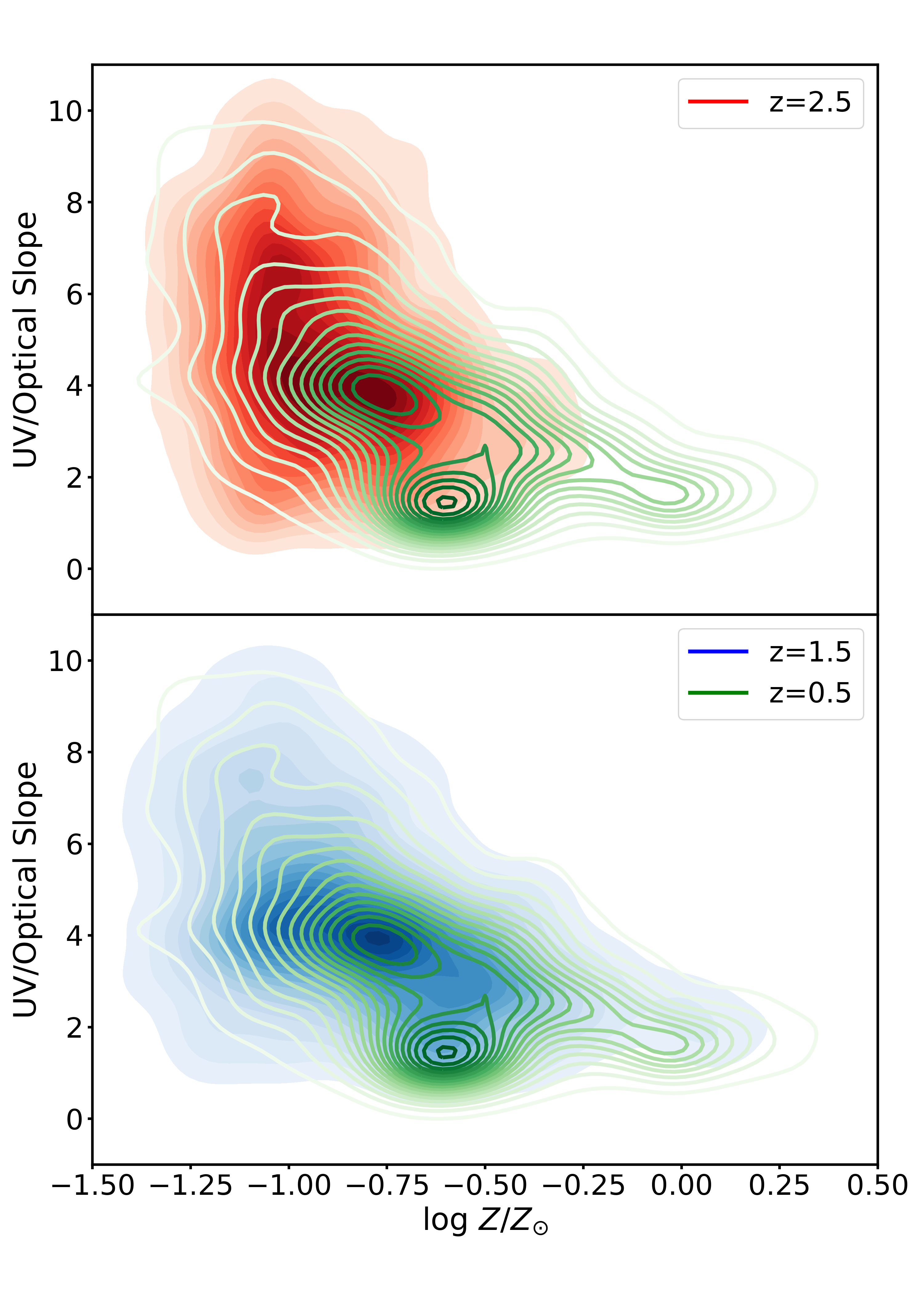}
    \caption{Contour plots of the UV-optical slopes $S$ (c.f. Equation~\ref{eq:slope}) of extinction curves against metallicities of the galaxies from our cosmological simulation. The red, blue and green contours represent galaxies at $z = 2.5$, $z=1.5$ and $z = 0.5$, respectively.  We show all galaxies to develop sufficient staistics (i.e. not just Milky Way analogs).   As in Figure~\ref{fig:ext_mw_evol}, as the metallicity increases at all redshifts, we see decreased slopes at these redshifts owing to highly efficient grain growth. \label{fig:ext_gal}}
\end{figure}

Understanding the origin of the Milky Way extinction law slopes and the relationship between bump strengths and slopes originates in the metallicity and dust growth history of the galaxy. 
To illustrate this, we first rewrite Equation~\ref{eq:bump} as $B \equiv (A_{\rm bump}/A_V) / (A_{\rm 2175,0}/A_V)$. $(A_{\rm 2175,0}/A_V)$ tends to increase as the slopes of the extinction curves become steeper. On the other hand, $(A_{\rm bump}/A_V)$ correlates with fraction of small graphite grains (we remind the reader of the extinction law shapes for pure graphites and pure silicates in Figure~\ref{fig:ext}).  We then show the relation between the slopes of the extinction curves and the gas-phase metallicities for all model galaxies (for better statistics) at $z \leq 2.5$ in Figure~\ref{fig:ext_gal}. There is an anti-correlation between the slopes and the metallicities, which is especially tight at $Z \gtrsim 0.3 Z_\odot$, due primarily to the highly efficient growth moving small grains (defined notionally here as $a \lesssim 0.06 \micron$) to the large grain regime ($a \gtrsim 0.06 \micron$)\footnote{We hereafter define the "Small to Large Ratio" (STL) as the mass fraction of grains smaller than $0.06 \micron$ cmopared to those larger than this notional size.}, and consequently flattening the extinction curves. This leads to decreasing bump strengths following flattening slopes as metallicities increase, provided the graphite to total dust mass ratios ($f_{\rm C}$) do not vary (which is the case for observed extinction curves along different lines of sight in the Milky Way). In reality, however, $f_{\rm C}$ span a large range even though STL stays the same. This results in substantial large scatter in our modeled $A_{\rm bump}/A_V$ relation, as well as  the resultant $B$--$S$ relation, with the upper bound corresponding to higher graphite to silicate ratio.

\begin{figure*}
    \centering
    \includegraphics[width=1.0\textwidth]{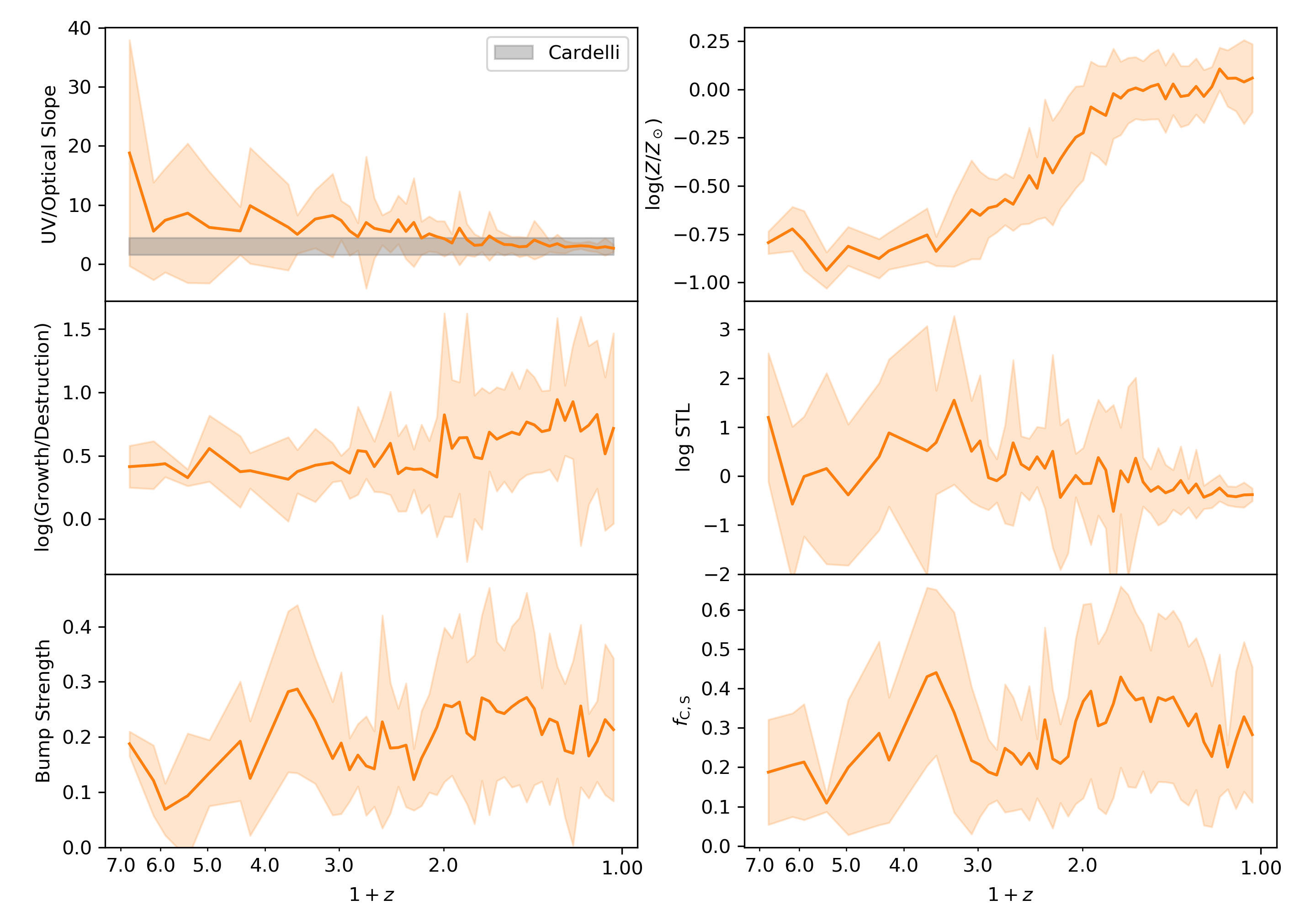}
    \caption{The slopes of extinction curves anticorrelate with metallicity in our model such that the slopes are reduced at late times/high metallicities.  This owes primarily to increased grain growth rates in high metal density environments. The bump strengths correlate with the graphite to total dust mass ratio such that a larger value leads to a bigger bump.  The subpanels here illustrate these trends, with details in \S~\ref{sec:ext}. Clockwise from top left: {\bf Top left:} We show the evolution of the UV/optical slope for $1$ Milky Way analog in our simulations (with the yellow shaded region showing dispersion for all progenitors of the same mass at a given redshift). The observed Milky Way range \citep{Cardelli1989} is shown by the light grey horizontal shaded region. {\bf Top right:} Evolution of the gas-phase metallicity. {\bf Middle right:} Evolution of the small grain to large grain ratio. {\bf Bottom right:} Evolution of graphite to total dust mass ratio.    {\bf Bottom left:} Evolution of the bump strength. {\bf Middle left:} Evolution of the ratio of grain growth to dust destruction rate. }
    \label{fig:ext_mw_evol}
\end{figure*}

In Figure~\ref{fig:ext_mw_evol}, we quantify the previous argument.  Here, we plot the redshift evolution of the (clockwise from top left) redshift evolution of the UV/optical slope of one of our model MW galaxies, gas phase metallicity, STL, the graphite to silicate ratio ($f_c$), the bump strength, and the grain growth to destruction rate rate.   To generate this plot, we follow the main progenitor of one of our MW analogs backwards in time via progenitors with the most number of stars in common (the solid orange lines).  To demonstrate the potential dispersion in this relation, at a fixed redshift we plot the dispersion in slopes and growth/destruction rates using galaxies whose stellar masses are within $0.3$ dex of the progenitor at that redshift.  For the redshift evolution of the extinction law slope, we show the observed Milky Way range \citep{Cardelli1989} with the light grey horizontal shaded region 

At early times while the metallicities are sufficiently low (redshift $z \gtrsim 2$), the masses of small grains double faster than the large grains, owing to their larger surface areas to volume ratios, and the extinction curves become rather steep.  As the galaxy enriches the ISM with metals, however, the metallicity-dependent grain growth rates (c.f. Equation~\ref{eq:3}) become sufficiently large to suppress the fraction of small grains, driving the small to large ratio (STL) down.  This increase in the relative fraction of large to small grains flattens the extinction law slopes to within the observed range of slopes.  

At the same time, the bump strengths depend primarily on the graphite to total dust mass ratio.  At early times, the silicates dominate as the dominant source of grain production is supernovae.  As AGB production becomes important ($z \lesssim 3$), however, the graphite fraction increases, and stabilizes (and hence, the bump strength does as well).  Stochastic variations in the graphite fraction at late times drive similar variations in the bump strength.  We now return to the bump-slope relation in Figure~\ref{fig:bump_slope}.  The variation in bump strengths at a fixed slope for our Milky Way analogs is  due to variations in the graphite to silicate ratio at $z=0$.  That there is any sense of a bump-slope relation in observations along Galactic sightlines likely reflects relatively constant graphite to silicate ratios within the Milky Way .

\section{Discussion}
\label{sec:discuss}
\subsection{Comparison to other Numerical Models}
\label{section:comparison}

While we reviewed the current status of the theoretical literature in this field in \S~\ref{section:introduction}, we remark briefly on numerical studies that specifically aim to understand the extinction law in Milky Way like galaxies here. 

\citet{Hou2019} adopted the $2$-size model of \citet{Hirashita2015} in order to investigate Milky Way extinction curves in cosmological simulations.  \citeauthor{Hou2019}'s simulation shows that the steepening of the extinction curves from $0.01<Z<0.2$ due to accretion and the flattening from $Z>0.2$ due to enhanced coagulation. In our simulation, this is mainly caused by the non-monotonic effect of grain growth processes on small-to-large grain mass ratios when the sources for the production of small grains are limited. This discrepancy mainly comes from the different treatment of grain growth and feedback. Grain growth in our simulation is overall stronger.  The difference may additionally owe to either the approximation of the continuous grain size distribution into two size bins by \citeauthor{Hou2019}, or alternatively to more subtle details of the {\sc simba} galaxy formation physics that we employ.

A more recent work employing similar methods has been performed by \cite{Huang2020}, who tracked a full spectrum of grain sizes in post-processed Illustris TNG model galaxies, though projected the galaxies into single zone models for computational efficiency.
 Their results shows a good match with an MRN grain size distribution at $z \sim 1$, though the extinction curves of the same Milky Way analogs grow steeper than the observed Galactic curve at lower redshifts (i.e. toward $z \sim 0$).  \citeauthor{Huang2020} argue that this is due to a slight drop of both galaxy metallicities and dense gas fractions (which is not observed in our simulations) that lead to reduced coagulation rates. It is also possible that grain growth processes, which we find to be important for setting the grain size distribution in our simulations, are not as effective as in our model.  Using one-zone models and fractions of dense gas computed by their equation $2$ to limit the growth rates could potentially lead to the underestimation of grain growth in star-forming dense regions where most dust mass exists. 

In an alternative class of models, \citet{McKinnon2018} and \citet{Aoyama2020} have developed on-the-fly models for a full spectrum of grain sizes in hydrodynamic simulations, as we have done here, though implemented these in idealized galaxy evolution models (i.e. those without a cosmological context). \citet{McKinnon2018} simulated the dust evolution in a Milky Way-mass disk galaxy in the absence of feedback, and indeed pioneered many of the equations and methods used in algorithms such as ours. \citeauthor{McKinnon2018} found extinction curves that were steeper than that of the Galaxy.  The overall effect of shattering on producing small grains in the \citeauthor{McKinnon2018} model could be too strong due to the lack of feedback that could enhance the destructive processes of dust and possibly their computation of grain velocities leading to a high relative speed of colliding grains. 
\citet{Aoyama2020} clarified the importance of resolution of the simulations, and showed the different grain size distributions and importantly developed physical insight as to how grain size distributions vary as a function of the physical properties of the ambient ISM.  

\subsection{Generalization of model predictions and caveats}
\label{section:caveats}
As we discussed in \S~\ref{section:introduction}, the observed extinction laws in the Magellanic clouds vary such that the mean curve of the LMC is steeper than that of the Galaxy, with reduced bump strength, and the SMC is steeper yet with no or very small UV bump.  This trend may be understandable from our simulations.


As is shown in Figure~\ref{fig:ext_gal}, our model predicts an overall anti-correlation between metallicities $Z$ and UV/optical slopes $S$ for galaxies at $z<2.5$, and the scatter is particularly tight at $Z> \sim 0.3 Z_\odot$. This trend is consistent with the fact that the Small Magellanic Cloud (SMC) and Large Magellanic Cloud (LMC) have st average metallicities in the SMC and LMC are $\sim0.2$ and $\sim 0.5$ $Z_\odot$, respectively \citep{Russell1992}.  This said, while we achieve extinction laws with a wide range of bump strengths and slopes (c.f. Figure~\ref{fig:bump_slope}), at metallicities comparable to the SMC and LMC, the vast majority of our simulated extinction curves have bump strengths larger than those observed in the Magellanic clouds.  This owes primarily to the fact that the bump strengths are dominated by the graphite to silicate ratio, which does not evolve as fast with metallicity as the STL ratio does. 


The lack of ability for our simulation to reproduce the location of the SMC in bump-slope space may represent an uncertainty in our simulation methods,  i.e. an  over-simplification when treating different grain species. Dust is assumed to be a mixture of graphite and silicate grains, but we do not evolve their grain size distribution separately, and omit the specific processes that may impact the lifecycle of 2175 \AA\ bump carriers (including UV-photon processing).   Beyond this, grain physics outside the scope of our current algorithms may contribute to the bump strength.  For example, small PAHs with sizes $a< 0.001 \micron$ may be associated with the UV bump  (e.g. \citealt{Mathis1994,Dwek1997,Li2001,Weingartner2001,Siebenmorgen2014}). Other models in the carbonaceous-silicate family include e.g amorphous carbons \citep{Zubko2004,Galliano2011} to replace non-PAH carbonaceous grains may help explain the bumpless feature in SMC environment (see  \citealt{Hirashita2020a}), provided that a detailed treatment specifically for the production and destruction of 2175\AA\ bump carriers (e.g. PAH) is included. Inclusion of the detailed treatment of multiple grain species will be a major future direction of our work in order to capture the full features of the extinction curves. 

\section{Conclusions}
\label{sec:conclude}
The main focus of this paper has been to understand the origin of, and variation in dust extinction curves in Milky Way-mass galaxies at $z \sim 0$.  To do this, we have developed a self-consistent model for evolving a distribution of dust grain sizes in cosmological hydrodynamic galaxy formation simulations that includes physical processes for dust formation in evolved stars, growth by the accretion of metals and coagulation, and destruction by thermal sputtering, grain shattering, and in star-forming regions.  We have confirmed that these models pass the benchmark of reproducing observed dust to gas ratio vs metallicity relations (Figure~\ref{figure:benchmark}).   Our main results follow:

\begin{enumerate}
\item Galaxies in our cosmological simulation with masses comparable to the Milky Way's exhibit a diverse range of modeled extinction laws, though they are all broadly within the range of curves observed within the Galaxy (Figure~\ref{fig:ext} and \S~\ref{sec:dsf}).  This broadly owes to modeled grain size distributions that converge to an MRN distribution at $z \sim 0$ in our simulations (Figure~\ref{fig:dmsf}). 

\item As Milky Way progenitors evolve from high redshift toward $z=0$, their extinction law slopes become flatter (to eventually be within the observed range of the Galaxy).  This owes to an increase in the ratio of the grain growth to destruction rates, which is a consequence of increased galaxy ISM metal densities at late times (Figure~\ref{fig:ext_mw_evol} and \S~\ref{sec:ext}). 

\item The bump strength is most closely dependent on the graphite to silicate ratio, which does not vary as strongly with the metallicity as the extinction law slope does.  At $z=0$, our model Milky Way analogs display both bump strengths and slopes comparable to the Milky Way (Figure~\ref{fig:bump_slope}), though do not demonstrate any clear relationship between the two owing to fluctuations in the graphite to silicate ratio (Figure~\ref{fig:ext_mw_evol}).  Whether a bona fide bump slope relationship in extinction laws exists is unclear from our simulations.


\item The increased extinction law slopes in our models at low metallicities may provide a natural explanation for the increased slopes in the LMC and SMC.  This said, because our model evolves the grain size distribution for graphites and silicates simultaneously, we are not able to reproduce the bumpless curves of the SMC on average (though some individual models do indeed exhibit similar bumpless and steep curves).  Future models that include models for evolving grain compositions are in progress.

\end{enumerate}

\section*{Acknowledgements}
Q.L. was funded by a graduate fellowship from the University of Florida Informatics Institute, as well as NSF AST-1909153. D.N. was funded in part by NSF AST-1715206 and AST-1909153.  The authors thank Hiroyuki Hirashita, Ryan McKinnon, Gerg\"o Popping, Samir Salim, Karin Sandstrom and J.D. Smith for helpful conversations.

\bibliographystyle{mnras}
\bibliography{references}
\onecolumn
\appendix

\section{Computing the Extinction Curve}
\label{appendix:extinction}
The optical depth at wavelength $\lambda$ contributed by grains with radii $a$ along a line of sight (LOS) is
\begin{equation}
    \tau (a,\lambda)\diff{a} = \int_{\rm LOS} \pi a^2 Q_{\rm ext} (a,\lambda) n_d(\mathbf{r},a)\diff{a}\ \diff{s},
\end{equation}
where $n_d(\mathbf{r},a)\diff{a}$ is the number density of grains with sizes $[a, a+\diff{a}]$ at location $\mathbf{r}$. The extinction efficiency $Q_{\rm ext} = Q_{\rm abs} + Q_{\rm sca}$ is the ratio of extinction to geometric cross section and considers effect of both absorption and scattering. Extinction efficiencies depend on the grain species (e.g. silicate or graphite in our simulation). Here we assume that the total carbon mass corresponds to the graphite mass, and the remainder goes to silicates. We adopt extinction efficiencies from \citet{Draine1984} and \citet{Laor1993}, interpolating their results to grain sizes we are interested in.

We can then get the extinction
\begin{equation}
    A(\lambda) = 2.5\log_{10} (e) \int_{a_{\rm min}}^{a_{\rm max}} \tau (a,\lambda)\diff{a}
               = 2.5\log_{10} (e) \int_{a_{\rm min}}^{a_{\rm max}} \diff{a}\ \pi a^2 Q_{\rm ext} (a,\lambda) \int_{\rm LOS} n_d(\mathbf{r},a)\diff{s}
\end{equation}

Considering the low resolution, we approximate $\int_{\rm LOS} n_d(\mathbf{r},a) \diff{s}$ by $\overline{n_d(a)}\ L$ where $\overline{n_d(a)}$ is the average number density of grains over the entire galaxies, the extinction curve $A(\lambda)/A(V)$ versus $1/\lambda$ is therefore generated by
\begin{equation}
    \left[ \frac{A(\lambda)}{A(V)} \right] = \frac{\int_{a_{\rm min}}^{a_{\rm max}} a^2 Q_{\rm ext} (a,\lambda) \overline{n_d(a)}\ \diff{a}}{\int_{a_{\rm min}}^{a_{\rm max}} a^2 Q_{\rm ext} (a,V) \overline{n_d(a)}\ \diff{a}}
\end{equation}

\end{document}